\documentstyle[pra,aps]{revtex}
\begin{document}
\draft

\def\overlay#1#2{\setbox0=\hbox{#1}\setbox1=\hbox to \wd0{\hss #2\hss}#1%
\hskip
-2\wd0\copy1}
\twocolumn[
\hsize\textwidth\columnwidth\hsize\csname@twocolumnfalse\endcsname

%\twocolumn[
\title{Ultrahigh sensitivity of slow--light gyroscope}
\author{U.\ Leonhardt$^{1,2}$ and P.\ Piwnicki$^2$}
\address{~$^1$School of Physics and Astronomy, University of St Andrews, 
North Haugh, St Andrews, Fife, KY16 9SS, Scotland}
\address{~$^2$Physics Department, Royal Institute of Technology (KTH),
Lindstedtsv\"agen 24, S-10044 Stockholm, Sweden}
\maketitle
%\mediumtext
\begin{abstract}
Slow light generated by Electromagnetically Induced Transparency is
extremely susceptible with respect to Doppler detuning.
Consequently, slow--light gyroscopes should have ultrahigh sensitivity.
\end{abstract}
\date{today}
%]
\vskip2pc]
\narrowtext
%\narrowtext
%%%%

In recent experiments \cite{Experiments} light has been slowed down
dramatically to just a few meters per second.
Traveling at this incredibly low speed, light is sensitive enough to
serve in detections of subtle motion effects such as the optical
Aharonov--Bohm effect of quantum fluids \cite{LPliten}.
A moving medium is able to drag light, and this dragging phenomenon
gives rise to the sensitivity with respect to motion.
Analyzed more carefully \cite{LPliten,LPstor,Gordon,PhamMauQuan},
a moving medium appears as an effective change of the space--time
metric (as an effective gravitational field).
When the medium outruns the light traveling inside, suitable flows
may appear as optical black holes \cite{LPliten}.
Apart from detecting quantum flows and from creating artificial black
holes, is there a practical application for the incredible
motion sensitivity of slow light?

Optical gyroscopes are regularly employed to sense motion.
What would be the advantage of a slow--light gyroscope?
How do passive dielectric gyroscopes work?
Imagine for simplicity a solid block of uniform dielectric material
that is rotating at angular velocity $\Omega_0$.
(In a real fiber--gyroscope light travels in a multitude of coils, 
for enhancing the effect.)
We assign cylindrical coordinates to the block where the $z$ direction
coincides with the rotation axis. 
First, assume an essentially non--dispersive material such as glass 
that is characterized by the refractive index $n$.
In this material, light experiences the space--time metric
\cite{LPstor}
\begin{equation}
\label{metric1}
ds^2 = \frac{c^2dt^2}{n^2} + 2\alpha\,\Omega_0\,r\,dt\,d\varphi
- dr^2 -r^2\,d\varphi^2 - dz^2
\end{equation}
in the limit of low rotation velocities $\Omega_0\,r$ compared with
the speed of light in vacuum, $c$.
Fresnel's dragging coefficient,
\begin{equation}
\label{fresneldrag}
\alpha = 1 - \frac{1}{n^2}
\,\,,
\end{equation}
quantifies the degree to which light is forced to move along with the
medium.
Clearly, $\alpha$ vanishes in the absence of a medium ($n=1$)
and $\alpha$ approaches unity in the limit of a very strong medium
($n\rightarrow\infty$).
Now, imagine that the block of dielectric material is at rest and
instead of the medium the laboratory frame is rotating at angular
velocity $\Omega$.
We obtain from the space--time metric of light in the medium frame,
\begin{equation}
\label{metric2}
ds^2 = \frac{c^2dt^2}{n^2} - dr^2 -r^2\,d\varphi'^2 - dz^2
\,\,,
\end{equation}
the metric in the rotating laboratory frame by the simple
transformation
\begin{equation}
\varphi' = \varphi - \Omega\,t
\,\,.
\end{equation}
To leading order in $\Omega r/c$ the transformed metric
(\ref{metric2}) coincides with the metric (\ref{metric1}) of light in
the rotating block if we put
\begin{equation}
\Omega = \alpha\,\Omega_0
\,\,.
\end{equation}
A rotating dielectric and a dielectric in a rotating frame are 
practically equivalent, yet they appear to rotate at different angular
velocities.
Fresnel's dragging coefficient (\ref{fresneldrag}) quantifies the
ratio between the actual and the apparent angular velocity of the
rotating dielectric body, $\Omega_0$ and $\Omega$, respectively.
The coefficient $\alpha$ characterizes the degree to which light 
follows the actual rotation and, consequently, the degree to which
rotation can be detected by optical interference.

Slow light \cite{Experiments} has been generated using
Electromagnetically Induced Transparency (EIT) \cite{EIT}.
EIT takes advantage of a quantum--interference effect in multi--level
atoms, bought about by dressing the atoms with the light of an
appropriate auxiliary beam. 
A probe beam at resonance frequency can travel though the EIT medium 
that would be totally opaque without the assistance of the auxiliary 
light.
Exactly on resonance a continuous probe wave travels at a phase
velocity of $c$, i.e. the medium has a refractive index of unity.
On the other hand, the EIT resonance is very sharp and is
ultrasensitive with respect to frequency detuning.
An EIT medium is thus extremely dispersive and, in turn, short light 
pulses having an extended spectrum travel at a very low group velocity
$v_g$. 
Another aspect of the extreme dispersion of an EIT medium is the 
ultrasensitivity with respect to the Doppler detuning due to motion.

So, what happens if we replace the rotating block of non--dispersive
material by an EIT medium of extreme dispersion?
How large is the dragging coefficient?
Let us determine the space--time line element,
\begin{equation}
ds^2 = g_{\mu\nu} dx^\mu dx^\nu
\quad,\quad
dx^\mu = \left(c\,dt,d{\bf x}\right)
\,\,,
\end{equation}
i.e. the covariant metric tensor $g_{\mu\nu}$.
For this we invert the contravariant metric of slow light
\cite{LPliten}, $g^{\mu\nu}$, to lowest order in the ratio between 
the medium velocity and the velocity of light, $c$.
In the limit of slow rotations, the line element turns out to have 
the same structure as the metric (\ref{metric1}) of non--dispersive 
media with, however, a refractive index of unity and a modified 
dragging coefficient of
\begin{equation}
\alpha = \frac{c}{v_g} - 1
\,\,.
\end{equation}
Compared with passive dielectric gyroscopes, a slow light gyroscope
operated with light of a modest group velocity of kilometers per
second increases the rotation sensitivity by five orders of
magnitude, and an ambitious group velocity of meters per second
amounts to a fantastic improvement by eight orders of magnitude.
Of course, for this one would need to create slow light in a solid
block of material in order to attach the gyroscope to the rotating
body one is interested in.
Solid--state media tend to destroy the quantum--interference
conditions of EIT much more rapidly than gases or Bose--Einstein
condensates \cite{Experiments}.
However, first demonstrations of EIT in solids have been already
reported \cite{EITsolids} and an interesting proposal of EIT in
semiconducters has been recently published \cite{Artoni}.
It would be desirable to demonstrate unambiguously slow light in
solids, stimulated perhaps by the potential advantage of slow light
gyroscopes: ultrahigh motion sensitivity.

\section*{Acknowledgements}

We are grateful to Malcolm Dunn and Stig Stenholm for valuable
discussions.
U.L. gratefully acknowledges the support of 
the Alexander von Humboldt Foundation
and of the G\"oran Gustafsson Stiftelse.

\end{document}